\documentclass[prd,superscriptaddress,twocolumn,nofootinbib,showpacs]{revtex4}
\usepackage{amsfonts,amssymb,amsmath}
\usepackage{color}
\usepackage{graphicx}
\usepackage{dcolumn}
\usepackage{bm}

\definecolor{refkey}{rgb}{0.9451,0.2706,0.4941}
\definecolor{labelkey}{rgb}{0.2706,0.4941,0.9451}

\newcommand{\C}[1]{{\mathcal #1}}

\newcommand{\beq}{\begin{equation}}
\newcommand{\eeq}{\end{equation}}
\newcommand{\bea}{\begin{eqnarray}}
\newcommand{\eea}{\end{eqnarray}}

\newcommand{\Tr}{\mathop{\rm Tr}}

\newcommand{\half}{\frac 12}
\newcommand{\third}{\frac 13}

\newcommand{\sixth}{\frac 16}

\newcommand{\Slash}[1]{{\ooalign{\hfil#1\hfil\crcr\raise.167ex\hbox{/}}}}

\begin{document}

\preprint{arXiv:1404.1450}

\title{TeV scale seesaw from supersymmetric Higgs-lepton inflation and BICEP2}
\author{Shinsuke Kawai}
\email{kawai(AT)skku.edu}
\affiliation{Department of Physics, 
Sungkyunkwan University,
Suwon 440-746, Republic of Korea}
\author{Nobuchika Okada}
\email{okadan(AT)ua.edu}
\affiliation{
Department of Physics and Astronomy, 
University of Alabama, 
Tuscaloosa, AL35487, USA} 

\date{5 April 2014}

\begin{abstract}
We discuss the physics resulting from the supersymmetric Higgs-lepton inflation model
and the recent CMB B-mode observation by the BICEP2 experiment.
The tensor-to-scalar ratio $r=0.20^{+0.07}_{-0.05}$
of the primordial fluctuations indicated by the CMB B-mode polarization is consistent with the prediction
of this inflationary model for natural parameter values.
A salient feature of the model is that it predicts the seesaw mass scale $M$ 
from the amplitude of the tensor mode fluctuations.
It is found that the 68\% (95\%) confidence level (CL) constraints from the BICEP2 experiment give
927 GeV $< M <$ 1.62 TeV
(751 GeV $< M <$ 2.37 TeV) for 50 e-foldings and 
391 GeV $< M <$ 795 GeV
(355 GeV $< M <$ 1.10 TeV) for 60 e-foldings.
In the type I seesaw case, the right-handed neutrinos in this mass range are elusive in collider experiments due to the small mixing angle.
In the type III seesaw, in contrast, the heavy leptons will be within the reach of future experiments.
We point out that a significant portion of the parameter region 
corresponding to the 68\% CL of the BICEP2 experiment
will be covered by the Large Hadron Collider experiments at 14 TeV.
\end{abstract}

\pacs{12.60.Jv, 
14.60.St, 
98.80.Cq, 
98.70.Vc
}

\keywords{Supergravity, right-handed neutrinos, inflation, cosmic microwave background}
\maketitle

\section{Introduction}
\label{sec:introduction}
The discovery of the cosmic microwave background (CMB) B-mode polarization by the BICEP2 experiment \cite{Ade:2014xna} is truly remarkable as the existence of the tensor mode in the primordial fluctuations provides a direct evidence for inflation in the early Universe\footnote{
The BICEP2 experiment uses 150 GHz single wavelength bolometers.
In order to conclude that the gravitational waves causing the polarization are undeniably of inflationary origin, the results need to be confirmed also at other wavelengths.}.
It has a significant impact on inflation model building.
In the past decade models producing small tensor mode fluctuations were considered
favourable since, for example, the Planck data in 2013 \cite{Ade:2013uln} constrained 
the tensor-to-scalar ratio $r < 0.11$ at 95\% confidence level (CL).
The models of inflation producing such small $r$ include the Higgs inflation model 
\cite{CervantesCota:1995tz,Bezrukov:2007ep}, supersymmetric Higgs inflation-type 
models \cite{Einhorn:2009bh,Ferrara:2010yw,Ferrara:2010in,Arai:2011nq,Arai:2013vaa}, 
the hill-top inflation model \cite{Boubekeur:2005zm}, and the $R^2$ inflation model 
\cite{Starobinsky:1980te}.
Among these, the Higgs inflation model is a particularly simple and concrete particle physics realization of inflation that also provides predictions in low-energy particle physics. 
These models are in tension with the finding of the BICEP2 experiment.
See Refs. \cite{Okada:2014lxa,Bezrukov:2014bra,Hamada:2014iga} for the updated status of 
various models.

In the present paper we point out that the prediction of the inflationary scenario
which we call the Higgs-lepton inflation (HLI) \cite{Arai:2011aa,Arai:2012em} fits extremely well with the new data for natural choice of parameters.
The HLI scenario is realized in the supersymmetric seesaw model, 
which is the simplest extension of the minimal supersymmetric Standard Model (MSSM) 
to include the right-handed neutrinos.
The model incorporates the type I \cite{seesaw} or type III seesaw mechanism \cite{Foot:1988aq} by which
the small nonzero neutrino masses that are evidenced by the neutrino oscillations are
naturally explained.
It also includes possibility for generating baryon asymmetry through leptogenesis or the Affleck-Dine
mechanism.
As a feature of the model, HLI directly associates the spectrum of the CMB with the
mass scale of the right-handed neutrinos.
We will see that the new data from the BICEP2 experiments constrains this mass scale
to be between a few hundred GeV and a few TeV.
These constraints are potentially useful since the right-handed (s)neutrinos may also be searched in
colliders.

\section{Higgs-lepton inflation in the supersymmetric seesaw model}
\label{sec:HLI}
The HLI model \cite{Arai:2011aa,Arai:2012em} is an ``all-in" phenomenological model 
of inflation that includes the seesaw mechanism \cite{seesaw}, the origin of the baryon asymmetry, the origin of the dark matter, as well as the Standard Model of particle physics.
It is based on the seesaw-extended MSSM.
The superpotential in the type I seesaw case is
\begin{align}
W=W_{\rm MSSM}+\half M N^c N^c+y_D N^c LH_u,
\label{eqn:W_1}
\end{align}
where $W_{\rm MSSM}$ is the MSSM superpotential and $N^c$, $L$, $H_u$ are the
right-handed neutrino singlet, the lepton doublet, and the up-type Higgs doublet superfields, respectively (the family indices are suppressed).
In the type III case, the superpotential is
\begin{align}
W=W_{\rm MSSM}+\half M\Tr\left(T^cT^c\right)+y_D LT^c H_u,
\end{align}
where
\begin{align}
T^c=\half\left(\begin{array}{cc}
N^0 & \sqrt 2 N^+\\
\sqrt 2 N^- & -N^0
\end{array}\right)
\label{eqn:W_3}
\end{align}
is the right-handed neutrino triplet superfield.
With odd R-parity assigned to $N^c$ or $T^c$, the superpotential preserves the R-parity in both cases.
The Majorana masses of the right-handed neutrinos $M$ and the Dirac Yukawa coupling $y_D$ are
related by the seesaw relation
\begin{align}
m_\nu=m_D^T M^{-1}m_D,
\label{eqn:seesawrel}
\end{align}
where $m_D=y_D\langle H_u^0\rangle$ and $\langle H_u^0\rangle\simeq 174$ GeV for moderate
$\tan\beta$.
While realistic seesaw requires at least two families of the 
right-handed neutrinos, we will be interested mainly in the outcome of inflation and
consider a simplified one family case\footnote{
See \cite{Arai:2012em} for a detailed description of the HLI with
two families (the minimal seesaw case) in type I seesaw.}.
Since the inflationary model is essentially the same for both type I and type III seesaw, we will describe in the case of the type I model below.
Estimating the mass scale of the light (left-handed) neutrinos as
$m_\nu^2\approx\Delta_{32}^2=2.32
\times 10^{-3}$ eV${}^2$ where the data of \cite{Beringer:1900zz} is used,
the seesaw relation \eqref{eqn:seesawrel} reads
\begin{align}
y_D=\left(
\frac{M}{6.29\times 10^{14} \mbox{ GeV}}\right)^{1/2}.
\label{eqn:yD}
\end{align}

Inflation is assumed to take place along one of the D-flat directions $L$-$H_u$, which is parametrized by a field $\varphi$ so that
\begin{align}
L=\frac{1}{\sqrt{2}}\left(\begin{array}{c}
\varphi\\0\end{array}\right),
\qquad
H_u=\frac{1}{\sqrt{2}}\left(\begin{array}{c}
0\\ \varphi\end{array}\right).
\end{align}
We consider supergravity embedding with slightly noncanonical
K\"{a}hler potential $K=-3\Phi$, where
\begin{align}
\Phi=1-\third \left(|N^c|^2+|\varphi|^2\right)
+\frac{\gamma}{4}\left(\varphi^2+ \mbox{c.c.}\right)
+\frac{\zeta}{3}|N^c|^4.
\label{eqn:Phi_1}
\end{align}
Here $\gamma$ and $\zeta$ are real parameters.
The third term on the right hand side violates the R-parity; the consequence of this
will be discussed in Section \ref{sec:TeV}.
We will use the unit in which the reduced Planck scale 
$M_{\rm P}=(8\pi G)^{-1/2}=2.44\times 10^{18}$ GeV is set to unity.
During inflation only the fields $N^c$ and $\varphi$ are important and the superpotential simplifies to
\begin{align}
W_{\rm inf}=\half MN^cN^c+\half y_DN^c\varphi^2.
\label{eqn:Winf}
\end{align}
From \eqref{eqn:Phi_1} and \eqref{eqn:Winf} the Lagrangian of the model can be
obtained following the standard supergravity computations
\cite{Arai:2011aa,Arai:2012em}.

The dynamics of the resulting system is complicated in general, with a nontrivial inflaton trajectory in multidimensional field space.
It can be shown however that with mild assumptions the model simplifies to give single-field slow roll inflation \cite{Arai:2011aa,Arai:2012em}.
This is due to the non-zero quartic term in \eqref{eqn:Phi_1}, which makes the
$N^c$ field massive, ensuring the inflaton trajectory to lie along the
$\varphi$ direction.
Furthermore, the scalar potential can be shown to be stable along the real axis of the $\varphi$
field so that the phase direction of $\varphi$ does not participate in the inflationary dynamics.
The model then involves only one real scalar field and the Lagrangian becomes
\begin{align}
{\C L}_{\rm J}=\sqrt{-g_{\rm J}}\left\{
\half\Phi R_{\rm J}-\half g_{\rm J}^{\mu\nu}\partial_\mu\chi\partial_\nu\chi-V_{\rm J}\right\},
\end{align}
where the subscript J stands for the Jordan frame and
\begin{align}
\chi=\sqrt 2\, {\rm Re}\,\varphi,
\qquad
V_{\rm J}=\frac{|y_D|^2}{16}\chi^4.
\end{align}
Here the field is understood to represent the scalar component.
Note that $\Phi$ of \eqref{eqn:Phi_1} is now written as
\begin{align}
\Phi=1+\xi\chi^2,
\qquad
\xi=\frac{\gamma}{4}-\frac{1}{6}.
\label{eqn:Phixi}
\end{align}
This is the nonminimally coupled $\lambda\phi^4$ model \cite{Okada:2010jf}.
The Higgs inflation model \cite{CervantesCota:1995tz,Bezrukov:2007ep} also 
has the same structure.
An essential feature of the HLI model here is that the inflaton self coupling is the square of the Yukawa coupling $y_D$ which is determined by the seesaw relation \eqref{eqn:seesawrel}.
In this supersymmetric model the effects of renormalization on the Yukawa coupling $y_D$ and 
the nonminimal curvature coupling $\xi$ are negligibly small \cite{Arai:2011aa,Arai:2012em}.

The dynamics of inflation and the prediction of the model can be studied 
conveniently in the Einstein frame, by Weyl-rescaling the metric
$g^{\rm E}_{\mu\nu}=\Phi g^{\rm J}_{\mu\nu}$ and
redefining the field $\chi$ into the canonically normalized one $\hat\chi$ in the Einstein frame,
\begin{align}
d\hat\chi=\frac{\sqrt{1+\xi\chi^2+6\xi^2\chi^2}}{1+\xi\chi^2}d\chi.
\end{align}
The Lagrangian in the Einstein frame is then
\begin{align}
{\C L}_{\rm E}=\sqrt{-g_{\rm E}}\left\{
\half R_{\rm E}-\half g_{\rm E}^{\mu\nu}\partial_\mu\hat\chi\partial_\nu\hat\chi
-V_{\rm E}\right\},
\end{align}
where the scalar potential is
\begin{align}
V_{\rm E}=\frac{V_{\rm J}}{\Phi^2}.
\end{align}
The slow roll parameters are defined in the usual way,
\begin{align}
\epsilon=\half\left(
\frac{1}{V_{\rm E}}\frac{dV_{\rm E}}{d\hat\chi}\right)^2,
\qquad
\eta=\frac{1}{V_{\rm E}}\frac{d^2V_{\rm E}}{d\hat\chi^2}.
\end{align}

\begin{figure}[t]
  \includegraphics[width=89mm]{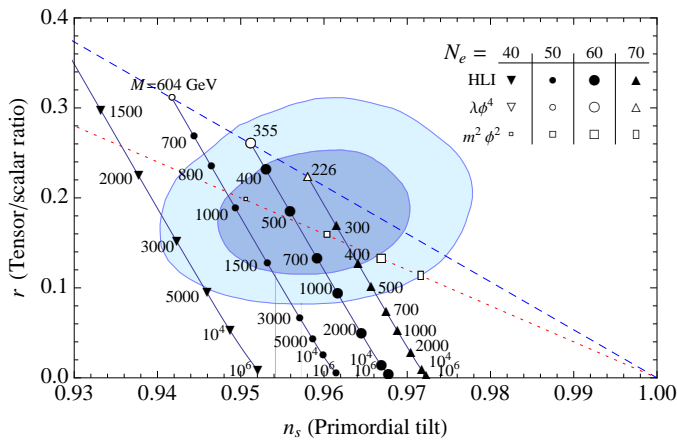}
  \caption{\label{fig:nsr_BICEP2}
  The spectral index $n_s$ and the tensor-to-scalar ratio $r$ of the Higgs-lepton inflation
  (HLI), for the e-foldings $N_e=40$, 50, 60 and 70. The nonminimal curvature coupling $\xi$ is fixed
  by the amplitude of the density fluctuations.
  The numbers shown alongside the plot are the seesaw scale $M$ measured in GeV.
  The 68\% and 95\% CL contours from the BICEP2 experiment \cite{Ade:2014xna} are shown 
  in the background.
  The prediction of the minimally coupled $m^2\phi^2$ chaotic model 
  [$n_s=1-2/(N_e+\half)$, $r=4(1-n_s)$, the red dotted line]
  and of the minimally coupled $\lambda\phi^4$ model
  [$n_s=1-3/(N_e+\frac 32)$, $r=\frac{16}{3} (1-n_s)$, the blue dashed line], 
  are also shown for comparison.}
\end{figure}

The model contains two tuneable parameters $y_D$ and $\xi$, which 
are related to $M$ and $\gamma$ through \eqref{eqn:yD} and \eqref{eqn:Phixi}.
The value of $\xi$ will be fixed by the amplitude of the CMB power spectrum
as follows.
The end of the slow roll is characterized by the condition that either of the slow roll parameters are not small anymore; we use ${\rm max}(\epsilon, |\eta|)=1$ and 
denote the value of the inflaton obtained from this condition as $\chi_*$.
We then follow the inflaton trajectory backward in time for $N_e$ e-foldings, 
and identify the inflaton value
$\chi_k$ that corresponds to the horizon exit of the comoving CMB scale $k$,
using the relation 
$N_e=\int_{\chi_*}^{\chi_k}d\chi V_{\rm E}
(d\hat\chi/d\chi)/(dV_{\rm E}/d\hat\chi)$.
Then the power spectrum of the curvature perturbation
$P_{R}=V_{\rm E}/24\pi^2\epsilon$ at the CMB scale $\chi=\chi_k$
is obtained for a given set of $N_e$, $y_D$, $\xi$.
To compare this with the observed CMB amplitude, we use for definiteness the value 
$A_s(k_0) = 2.215\times 10^{-9}$ from the Planck satellite experiment \cite{Ade:2013uln}, 
with the pivot scale at $k_0=0.05$ Mpc${}^{-1}$.
Here, $A_s(k)=\frac{k^3}{2\pi^2}P_R(k)$ and $P_{R}(k)$ is the Fourier transform of 
$P_{R}$.
Fixing $\xi$ by this procedure we obtain the prediction of the CMB spectrum for a given number of 
e-foldings $N_e$ and a value of the Yukawa coupling $y_D$.
The prediction for the scalar spectral index $n_s
=1-6\epsilon+2\eta$ and the tensor-to-scalar ratio $r\equiv P_{\rm gw}/P_{R}=16\epsilon$ are plotted in Fig.\ref{fig:nsr_BICEP2}. 
Instead of the Yukawa coupling $y_D$, the seesaw scale $M$ is shown in the figure.
In the background we also indicate the 68\% and 95\% CL constraint contours 
from the BICEP2 experiment \cite{Ade:2014xna}.

An important feature of the HLI model is that the inflaton quartic self coupling is given by the
square of the Dirac Yukawa coupling, which in turn is related to the mass scale of the right-handed
(s)neutrinos by the seesaw relation \eqref{eqn:yD}.
There is a lower bound of the Yukawa coupling, which is determined in the minimal coupling limit 
$\xi\to 0$ by the CMB amplitude.
One can see from Fig.\ref{fig:nsr_BICEP2} that the constraints from the BICEP2 experiment yield
the seesaw mass scale in the range between a few hundred GeV and a few TeV.
More concretely we find
\begin{align}
927\; {\rm GeV} &< M < 1.62\; {\rm TeV}& (\mbox{68\% CL})\cr 
751\; {\rm GeV} &< M < 2.37\; {\rm TeV}& (\mbox{95\% CL})
\label{eqn:MforN50}
\end{align}
for $N_e=50$
and 
\begin{align}
391\; {\rm GeV} &< M < 795\; {\rm GeV}& (\mbox{68\% CL})\cr
355\; {\rm GeV} &< M < 1.10\; {\rm TeV}&(\mbox{95\% CL})
\label{eqn:MforN60}
\end{align}
for $N_e=60$.
In Table \ref{table:table1} the values of the Dirac Yukawa coupling $|y_D|$, the nonminimal curvature
coupling $\xi$, the scalar spectral index $n_s$ and the tensor-to-scalar ratio $r$ for these 68\% and 95\% threshold cases are shown.
Note that the small inflaton quartic coupling $\sim 10^{-12}$ is not unnaturally small,
since it is the {\em square} of the Dirac Yukawa coupling.
In fact the Dirac Yukawa coupling of $y_D\sim 10^{-6}$ is in the same order as the electron Yukawa coupling $y_e$.

\begin{table}[t]
\begin{center}\begin{tabular}{c|ccccc}$N_e$ & $M$ (GeV) & $|y_D|$ & $\xi$ & $n_s$ & $r$ \\ 
\hline\\
& 751 & $1.09\times 10^{-6}$ & $5.89\times 10^{-4}$ & 0.946 & 0.250 \\
50 & 927 & $1.21\times 10^{-6}$ & $1.29\times 10^{-3}$ & 0.948 & 0.203 \\ 
& $1.63\times 10^3$ & $1.61\times 10^{-6}$ & $4.00\times 10^{-3}$ & 0.954 & 0.118 \\ 
& $2.37\times 10^3$ & $1.94\times 10^{-6}$ & $6.86\times 10^{-3}$ & 0.956 & 0.0822 \\ 
\hline\\
& 355 & $7.52\times 10^{-7}$ & 0 & 0.951 & 0.260 \\
60 & 391 & $7.89\times 10^{-7}$ & $2.09\times 10^{-4}$ & 0.953 & 0.236 \\ 
& 795 & $1.12\times 10^{-6}$ & $2.49\times 10^{-3}$ & 0.960 & 0.117 \\ 
& $1.10\times 10^3$ ~ & $1.32\times 10^{-6}$ ~ & $4.19\times 10^{-3}$ ~ & 0.962 ~ & 0.0855
\end{tabular} 
\caption{
\label{table:table1}
The values of the Yukawa coupling $|y_D|$, the nonminimal curvature coupling $\xi$, the scalar
spectral index $n_s$ and the tensor-to-scalar ratio $r$ of the HLI model with e-foldings 50 and 60.
The chosen seesaw scales $M$ correspond to the 68\% and 95\% CL contours of the BICEP2
experiment.}
\end{center}
\end{table}

\section{Physics implied by the TeV seesaw scale}
\label{sec:TeV}

We now turn to discuss various features of the HLI scenario when the seesaw scale is in the range \eqref{eqn:MforN50}, \eqref{eqn:MforN60}.

\subsection{Reheating temperature}

The inflaton $\varphi$ of the HLI model is the $L$-$H_u$ flat direction of the supersymmetric seesaw model.
The dominant decay channel of the Higgs component is $\varphi\to b\bar b$.
In the perturbative reheating scenario, the upper bound of the reheating temperature is then
estimated using the decay rate as $T_{\rm rh}\lesssim 10^{7}$ GeV.
Parametric resonance effects and/or contributions from other decay channels may slightly alter
this estimate.
Allowing for the redshift before the Universe reaches thermalization, we evaluate the
reheating temperature of this model to be 
\begin{align}
T_{\rm rh}\approx 10^5 - 10^7 \mbox{ GeV}.
\label{eqn:Trh}
\end{align}
The effects of the nonminimal coupling should be negligible, since such effects become important
only when $\xi$ is extremely large and the coupling of the inflaton with the particle to which it decays is extremely small \cite{Bassett:1997az}.

\subsection{Baryon asymmetry}

The reheating temperature \eqref{eqn:Trh} is lower than the grand unification scale.
Thus, the baryon asymmetry of the Universe (BAU) needs to be produced in a mechanism other than in the
GUT phase transition.
The supersymmetric seesaw model includes the right-handed (s)neutrinos and there is possibility
that the BAU can be produced by the leptogenesis scenario \cite{Fukugita:1986hr,Albright:2003xb},
in which the lepton number first generated by the out-of-equilibrium decay of the right-handed
(s)neutrinos is later converted into the baryon number via $(B+L)$-violating sphaleron transitions.
We found in Refs \cite{Arai:2011aa,Arai:2012em} that when the seesaw scale is higher the leptogenesis
scenario operates successfully.
Let us see how the scenario is altered when the seesaw mass scale is \eqref{eqn:MforN50}, \eqref{eqn:MforN60} 
conforming to the results of the BICEP2 experiment.

As the masses of the right-handed (s)neutrinos are much smaller than the reheating temperature
\eqref{eqn:Trh}, the right-handed (s)neutrinos are thermally produced in the reheating process.
The question of whether the thermal leptogenesis operates well or not may be studied in two steps:
(i) whether sufficient lepton asymmetry is generated by the decay of the (s)neutrinos, and (ii)
whether the lepton number is successfully converted into the observed abundance of the baryon number, namely $Y_B\equiv n_B/s\sim 10^{-10}$ where $n_B$ is the baryon number density and $s$ is the entropy density.

In the traditional thermal leptogenesis with hierarchical seesaw masses, generation of sufficient
lepton asymmetry requires large enough seesaw mass $M\gtrsim 6\times 10^8$ GeV 
\cite{Davidson:2002qv} and high enough reheating temperature $T_{\rm rh}\gtrsim 10^9$ GeV
(see e.g. \cite{Buchmuller:2005eh}).
These conditions are based on assumptions such as the hierarchy of the seesaw masses and the 
flavor structure, and may be relaxed.
In particular, it is known that when at least two of the right-handed neutrino masses are nearly degenerate, resonance enhancement of the $CP$-asymmetry parameter takes place, resulting in 
larger BAU for lower reheating temperature (resonant leptogenesis
\cite{Flanz:1996fb,Pilaftsis:1997jf,Pilaftsis:2003gt}).
In fact, for $M\gtrsim$ TeV in the case of the minimal type I seesaw case (with two families of $N^c$), 
we have shown in \cite{Arai:2012em} that sufficient BAU can be produced in the HLI model, both for the 
normal mass hierarchy and for the inverted mass hierarchy.
However, when $M$ is below a few TeV, the efficiency of the sphaleron process is strongly suppressed,
diminishing the conversion of the lepton number into the baryon number.
In the type III seesaw the lower bound of the seesaw mass for successful leptogenesis is 
\cite{Strumia:2008cf,Hambye:2012fh}
\begin{align} 
M > 1.6 \;\mbox{TeV}.
\end{align}
The lower bound of the type I seesaw mass is in the same order (see e.g. \cite{Iso:2010mv}).

To conclude, the seesaw mass scale \eqref{eqn:MforN50}, \eqref{eqn:MforN60} is too small for
successful thermal leptogenesis, as the sphaleron process becomes inefficient and the lepton number
cannot be successfully converted into the baryon number.
The HLI model is equipped with supersymmetry and the BAU may be generated for example by the Affleck-Dine mechanism \cite{Affleck:1984fy}.

\subsection{R-parity violation}

The third term of the K\"{a}hler potential \eqref{eqn:Phi_1} is proportional to $\gamma LH_u+c.c.$,
which violates the R-parity\footnote{
Small R-parity violating terms are employed to solve various problems.
See e.g. \cite{Dreiner:1997uz}.}.
For the small values of the nonminimal coupling $\xi$ (see Table \ref{table:table1}), 
the coefficient of this term is $\gamma=4(\xi+\sixth)\approx {\C O}(1)$.
Along with supersymmetry breaking, this term induces an R-parity violating effective 
superpotential of the form $W_{\scriptsize\Slash{R}}\sim\mu' LH_u$, where 
$\mu'=\gamma F_\phi^\dag$ and $F_\phi$ is the conformal compensator F-term.
Combined with the usual MSSM $\mu$-term, the superpotential of the form
$W\sim H_u(\mu H_d+\mu'L)$ generates lepton number violating terms
\begin{align}
W_{\Delta L=1}\sim (y_e\varepsilon)e^c LL+(y_d\varepsilon) d^c QL,
\label{eqn:WL}
\end{align}
with $\varepsilon\sim\mu'/\mu$.
These terms are potentially hazardous as they hamper baryogenesis (which we assume to take place
by the Affleck-Dine mechanism).

In generic models of supersymmetry breaking, the gravitino mass is given by 
$m_{3/2}\approx F_\phi$ and hence $\mu'\approx m_{3/2}$.
The cosmological constraints on the size of the effective Yukawa couplings in \eqref{eqn:WL}
indicate
$\varepsilon\sim m_{3/2}/\mu\lesssim 10^{-6}$ 
\cite{Campbell:1990fa,Fischler:1990gn,Dreiner:1991pe}.
R-parity violating terms may also generate neutrino masses (separately from the seesaw mechanism) and
the condition that such effects do not lead to unacceptably large neutrino masses gives somewhat weaker constraint $\varepsilon\lesssim 10^{-3}$ \cite{Hempfling:1995wj,Nilles:1996ij,Hirsch:2000ef}.
Using the typical value of the MSSM $\mu$ parameter $\mu\sim 1$ TeV, 
$\varepsilon\lesssim 10^{-6}$ gives the upper bound of the gravitino mass
$m_{3/2}\lesssim 1$ MeV.
In practice, the consequence of the R-parity violation largely depends on the details of 
assumed scenario of supersymmetry breaking \cite{Arai:2012em}.
For example, in the "almost no-scale" scenario \cite{Luty:2002hj}, we have $m_{3/2}\gg F_\phi$ and
consequently the gravitino mass can be much larger than 1 MeV.

\subsection{Dark matter candidates}

As the gravitino mass of $\sim 1$ MeV is much smaller than the typical neutralino or slepton masses,
the lightest superparticle (LSP) of the HLI model is the gravitinos.
It also can be shown that for the small R-parity violation the gravitinos are sufficiently long-lived, and
the mass of $\sim 1$ MeV is large enough for cold dark matter particles.
Thus the gravitinos are a good candidate of the dark matter in this inflationary scenario. 

The overproduction constraints of the thermally produced gravitinos give further restriction
on the reheating temperature.
The abundance of the gravitinos is \cite{Bolz:2000fu,Pradler:2006qh,Steffen:2006hw}
\begin{align}
\Omega_{3/2} h^2\simeq 0.3\times\left(\frac{T_{\rm rh}}{10^{10} \mbox{ GeV}}\right)
\left(\frac{100 \mbox{ GeV}}{m_{3/2}}\right)
\left(\frac{M_{\tilde g}}{1 \mbox{ TeV}}\right)^2,
\end{align}
where $M_{\tilde g}$ is the running gluino mass and $h\approx 0.670$ is the Hubble parameter measured in 100 km Mpc${}^{-1}$ sec${}^{-1}$.
Now using $m_{3/2}\approx 1$ MeV, $M_{\tilde g}\approx 1$ TeV and $\Omega_{3/2}\approx 0.1$, 
the reheating temperature is found to be $T_{\rm rh}\approx 10^5$ GeV.
This is consistent with the reheating temperature estimated previously in \eqref{eqn:Trh}.

\subsection{Collider physics}

Finally, let us comment on implication of our inflationary scenario in collider physics.
Candidates of the seesaw particles are actively searched e.g. in the Large Hadron Collider (LHC) 
\cite{Bajc:2006ia,Bajc:2007zf,Franceschini:2008pz,Arhrib:2009mz,CMS:2012ra,ATLAS:2013hma}.
It would be important to discuss detectability of the seesaw particles in the mass range 
\eqref{eqn:MforN50}, \eqref{eqn:MforN60} .
In the case of type I seesaw, production of the singlet neutrinos in colliders is due to mixing with the
doublet neutrinos. 
However, the smallness of the Yukawa coupling corresponding to the TeV scale seesaw indicates that the mixing angle is too small, and production in colliders is negligible 
\cite{Gonzalez:2013ufa,Atre:2009rg}.
In the type III seesaw case, in contrast, the triplet fermions are produced by the electroweak gauge
interactions, and production in LHC is in principle possible.
Indeed, the type III seesaw particles of mass scale below 245 GeV have already been excluded at 95\% CL
by the ATLAS experiment \cite{ATLAS:2013hma} (CMS gives similar lower bound 180-210 GeV
\cite{CMS:2012ra}).
The coverage of the LHC at 14 TeV run is expected to be up to 750 GeV \cite{delAguila:2008cj}.
This means that in Fig.\ref{fig:nsr_BICEP2} significant part of the parameter space inside the
68\% contour will be covered.
It would be interesting to see whether the type III HLI model survives this test.

\section{Summary and Discussion}
\label{sec:concl}

We have discussed in this paper implications of the BICEP2 results in the
Higgs-lepton inflation (HLI) model.
The large nonzero value of the tensor mode $r=0.20^{+0.07}_{-0.05}$ calls for 
significant change of its interpretation.
We have found that the HLI model can be fit with the new CMB data, it stays consistent with the
neutrino oscillation data, and the thermally produced gravitinos remain a viable candidate of the
cold dark matter.
The BAU, nevertheless, cannot be generated by the leptogenesis scenario; due to the smallness
of the seesaw mass scale, the sphaleron process becomes inefficient and the
lepton asymmetry generated along with the right-handed (s)neutrinos cannot be converted into 
baryon asymmetry.
We discussed that the Affleck-Dine mechanism is a possible scenario of baryogenesis.
We also pointed out that in the type III seesaw case, the triplet fermions in the mass range conforming
to the BICEP2 experiment can be searched in the LHC at 14 TeV run.

One of the original motivations for the HLI model \cite{Arai:2011aa,Arai:2012em}
was to overcome the shortcomings of the Higgs inflation model \cite{CervantesCota:1995tz,Bezrukov:2007ep} which predicts small tensor mode fluctuations at the cost of introducing extremely large
nonminimal curvature coupling $\xi\sim 10^4$.
Due to the discovery of the large tensor mode fluctuations of inflationary origin by the BICEP2 experiment, the Higgs inflation model lost its attractiveness and the good-old $\lambda\phi^4$ and $m^2\phi^2$ minimally coupled chaotic inflation models have resurfaced as favored models.
The HLI model with small seesaw mass scale which we discussed in the present paper is in fact
an {\em almost} minimally coupled $\lambda\phi^4$ model.
Let us point out however that the HLI model has at least two advantages over the $\lambda\phi^4$
chaotic inflation model.
One is that it is a concrete particle physics realization of inflation with predictive power in particle phenomenology.
The other advantage is that the extremely small inflaton self coupling $\lambda\sim 10^{-12}$ is
not unnatural in HLI, since $\lambda\sim |y_D|^2$ and $|y_D|\sim 10^{-6}$ which is, while small, in the same order 
as the electron Yukawa coupling.

In view of the BICEP2 results, the $m^2\phi^2$ chaotic inflation model driven by the right-handed scalar neutrinos \cite{Murayama:1992ua} is an attractive model, sharing many aspects with the one presented in this paper.
Both inflationary models are realized in the supersymmetric seesaw model.
They arise as different choices of K\"{a}hler potential and inflaton trajectory.
In their model \cite{Murayama:1992ua}, leptogenesis is always successful and the seesaw scale is necessarily large, $M\sim 10^{13}$ GeV.
Seesaw particles of such large masses are, unlike in our model here, obviously far beyond the reach of any collider experiments.
It would be interesting if the future CMB or other observations distinguish these models.

{\em Acknowledgments.} --- 
This work was supported in part by the National Research Foundation of Korea 
Grant-in-Aid for Scientific Research 
No. 2013028565 (S.K.) and by the DOE Grant No. DE-FG02-10ER41714 (N.O.).
We used computing resources of the Yukawa Institute, Kyoto University.


\end{document}